\documentclass[preprintnumbers,amsmath,amssymb,floatfix,10pt,prd,twocolumn,superscriptaddress,nofootinbib]{revtex4-2}
\usepackage{latexsym}
\usepackage{epsfig}
\usepackage{epstopdf}
\usepackage{graphicx}
\usepackage{amssymb}
\usepackage{amsmath}
\usepackage{amsfonts}
\usepackage{subfigure}
\usepackage{dcolumn}
\usepackage{bm}
\usepackage{color}
\usepackage{comment}
\usepackage[utf8]{inputenc}
\usepackage[dvipsnames]{xcolor}
\usepackage{hyperref}
\hypersetup{colorlinks=true,linkcolor=blue,citecolor=red,urlcolor=magenta}
\usepackage{orcidlink}

\begin{document}
\title{\bf Influence of Quantum Correction on Kerr Black Hole in Effective Loop Quantum Gravity via Shadows and EHT Results}

\author{Muhammad Ali Raza \orcidlink{0009-0001-1281-8702}}
\email{maliraza01234@gmail.com}
\affiliation{Department of Mathematics, COMSATS University Islamabad, Lahore Campus, Lahore, Pakistan}

\author{M. Zubair}
\email{mzubairkk@gmail.com;drmzubair@cuilahore.edu.pk}
\affiliation{Department of Mathematics, COMSATS University Islamabad, Lahore Campus, Lahore, Pakistan}
\affiliation{National Astronomical Observatories, Chinese Academy of Sciences, Beijing 100101, China}

\author{Farruh Atamurotov}
\email{atamurotov@yahoo.com}
\affiliation{Urgench State University, Kh. Alimdjan str. 14, Urgench 220100, Uzbekistan}
\affiliation{University of Tashkent for Applied Sciences, Str. Gavhar 1, Tashkent 100149, Uzbekistan}

\author{Ahmadjon Abdujabbarov}
\email{ahmadjon@astrin.uz}
\affiliation{New Uzbekistan University, Movarounnahr street 1, Tashkent 100000, Uzbekistan}
\affiliation{Shahrisabz State Pedagogical Institute, Shahrisabz Str. 10, Shahrisabz 181301, Uzbekistan}

\begin{abstract}
Recently, a study on shadow of quantum corrected Schwarzschild black hole in loop quantum gravity appeared in [Ye et al., \textit{Phys. Lett. B} 851, 138566, (2024)] assuming a fixed value of Barbero-Immirzi parameter $\gamma$. Following this approach, we considered its rotating counterpart being a quantum corrected Kerr black hole in effective loop quantum gravity and studied its deviation from Kerr black hole for a fixed value of $\gamma$. We proposed and proved a theorem describing the location of unstable circular null orbits for all such Kerr-like metrics. The deviation between the shadows of the Kerr and quantum corrected Kerr black holes has also been studied, and parameters are constrained by comparison with the EHT results for M87* and Sgr A* to precisely probe the quantity of deviation due to quantum correction. Lastly, we immersed the quantum corrected Kerr black hole in an inhomogeneous plasma and studied its impact on the shadow size. We found that the unstable null orbits for the quantum corrected Kerr black hole are always smaller than the unstable null orbits for Kerr black hole. The effect of Barbero-Immirzi parameter allows the quantum corrected Kerr black hole to mimic Sgr A* with a higher probability than the Kerr black hole. However, the quantum corrected Kerr black hole does not mimic M87*. The plasma reduces the size of the shadow of quantum corrected black hole, and the plasma parameter in the case II is more sensitive than that in case I.
\end{abstract}
\maketitle
\date{\today}

\section{Introduction}\label{S1}
Soon after the foundation of General Relativity (GR), Schwarzschild \cite{1916SPAW.......189S} proposed a black hole (BH) metric as the pioneering solution of Einstein field equations in vacuum. However, this BH metric encompasses a spacetime singularity, a region where all physical laws breakdown and nothing can be explained \cite{2000Prama..55..529J}. Though, in classical regime, GR is a well understood theory and may explain various phenomena in gravitational physics. However, it has failed to resolve the singularity problem. Penrose \cite{PhysRevLett.14.57} proposed that the spacetime singularity is inevitable, and later, Hawking and Penrose \cite{1970RSPSA.314..529H} proposed the inevitability of the singularity in Big Bang.

To resolve the singularity issue, various attempts have been made over the years and thus one may expect a quantum theory of gravity may resolve the issue. One of the proposed theories for quantum gravity is loop quantum gravity (LQG), characterized by its independence from a fixed background and its non-perturbative approach \cite{Rovelli_2004,Thiemann_2007,2004CQGra..21R..53A,2007IJMPD..16.1397H,2013LRR....16....3P}. The theoretical and numerical aspects of Loop Quantum Cosmology (LQC) have provided solutions to the cosmological big-bang singularity \cite{PhysRevLett.86.5227,2003gr.qc.....4074A,PhysRevLett.96.141301,PhysRevD.73.124038,PhysRevD.74.084003}. Some approaches to resolve the singularity of the Schwarzschild BH involve quantization of its interior using techniques derived from LQG \cite{2006CQGra..23..391A,PhysRevD.76.104030,2010IJTP...49.1649M,2016CQGra..33e5006C,PhysRevLett.121.241301,PhysRevD.98.126003,PhysRevD.102.041502,PhysRevD.105.024069}. Moreover, studies have explored the LQG corrections related to BH formation and gravitational collapse in different theoretical models \cite{PhysRevLett.95.091302,PhysRevD.80.084002,PhysRevD.104.046019,PhysRevLett.128.121301,PhysRevD.104.106017}.

Recently, Lewandowski et al. \cite{PhysRevLett.130.101501} investigated gravitational collapse of a dust ball by incorporating quantum effects from LQG, using a LQC framework. They found that the collapse halts when the dust ball's energy density reaches the Planck scale, causing the dust to bounce instead of continuing to collapse. By matching the metrics at the boundary between the collapsing dust ball's interior and exterior, they derived a quantum correced Schwarzschild metric for the external spacetime. This work highlights how quantum gravity influences the behavior of collapsing structures. Later, Ye et al. \cite{YE2024138566} studied the shadow and photon rings of this quantum corrected BH. He found that this quantum corrected BH can be distinguished from the Schwarzschild BH in terms of shadow images in some illumination models. They assumed the fixed value of the Barbero-Immirzi parameter and hence the parameter $\alpha$ was not treated as a free parameter. Their findings were presented in comparison with the results of Schwarzschild BH. Motivated by this, assuming the fixed value of the Barbero-Immirzi parameter, we will accomplish our analysis in comparison with the results of Kerr BH for the rotating case.

In 2000, Falcke et al. \cite{2000ApJ...528L..13F} predicted the possibility of imaging the BHs especially Sgr A*, which got verified by the discovery of images of the supermassive BH M87* \cite{EHT2019,EHT2019b} and Sgr A* \cite{EHT2022,EHT2022b} by the Event Horizon Telescope (EHT) that has opened up new avenues in BH physics. These images comprise light rings and shadows of the BHs with nearly equal radii. A light ring is formed by the trapping of light and appears as a glowing image of radius greater than the event horizon. Whereas, the shadow is not a physical entity, but a dark 2D silhouette formed by the disappearing of photons from the sight of an observer at infinity. Since then, many studies have been accomplished to understand the shape and size of the BH shadow in various different models and frameworks, see Refs. \cite{PhysRevD.85.064019,PhysRevD.87.044057,PhysRevD.88.064004,PhysRevD.89.124004,PhysRevD.90.024073,2016EPJC...76..273A,PhysRevD.95.104058,2018GReGr..50...42C,2018NatAs...2..585M,PAPNOI2022100916,Atamurotov_2022,2022EPJC...82..771S,2022EPJC...82..831A}. The images of M87* and Sgr A* provided the data which has been useful in testing various gravity theories by determining constraints on the parameters associated with the theories, see Refs. \cite{Afrin_2022,2022EPJC...82.1018J,PANTIG2023169197,Hendi_2023,Islam_2023,Afrin_2023,2023EPJC...83..250P,UNIYAL2023101178,2023EPJC...83..588N,DAVLATALIEV2023101340,JAFARZADE2024101497,PhysRevD.100.044057,PhysRevD.104.024001,Vagnozzi_2022,PhysRevD.106.084051,Sengo_2023,Vagnozzi_2023,ATAMUROTOV2024101625}. In particular, Islam et al. \cite{Islam_2023} investigated LQG by using the EHT data for M87* and Sgr A*. They considered a rotating polymerized BHs in LQG which acts as Kerr BH asymptotically. Using the shadow analysis, the LQG parameter has been constrained and found that a significant part of parametric spaces for one and two horizon BHs is consistent with the EHT results for both M87* and Sgr A*. Moreover, the EHT results for Sgr A* also agree with the triple horizon BH, but not for the M87*. Afrin et al. \cite{Afrin_2023} investigated the LQG for the results of M87* and Sgr A* by considering two rotating LQG-inspired BHs and found that the upper bound for LQG parameter obtained from the results of Sgr A* is more precise than the upper bound from M87*. The astrophysical BHs are surrounded by plasma due to a huge gravity and in-falling matter \cite{EHT2022bc}. This may certainly last an impact on the light propagation causing deviation in shadows. The study of light propagation in magnetized plasma with no pressure, is governed by the Hamiltonian derived by Breuer and Ehlers \cite{doi:10.1098/rspa.1980.0040,doi:10.1098/rspa.1981.0011}. Later, Perlick and Tsupko \cite{PhysRevD.95.104003} derived the Hamiltonian formulation for a rather simpler case of a plasma without pressure and magnetization. Perlick et al. \cite{PhysRevD.92.104031} also explored the shadow of static and spherically symmetric BHs immersed in a plasma medium. Recently, various BH solutions have been considered to investigate their shadows in the presence of plasma, see Refs. \cite{PhysRevD.104.064039,PhysRevD.107.024027,PhysRevD.107.124004,PhysRevD.92.084005,Das_2022,DAVLATALIEV2023101340,Raza_2024}.

Inspired by \cite{YE2024138566}, we consider the recently developed rotating counterpart \cite{2024arXiv240917099F,2024arXiv241009198A,2024arXiv241011332V} of the quantum corrected Schwarzschild BH \cite{PhysRevLett.130.101501}, since the supermassive BHs are rotating in nature. Our major goal is to test the LQG effects under the influence of Barbero-Immirzi parameter, by comparing its shadow results with the EHT data and for Kerr BH. In particular, we investigate the deviation of the quantum corrected Kerr BH from the Kerr BH via EHT results. We also study the impact of plasma on the shadow of quantum corrected Kerr BH. The paper is organized as: In Sect.~\ref{S2}, we present a brief overview of the static and rotating BH metrics, and further discuss its horizon structure in terms of BH spin. In Sect.~\ref{S3}, we employ the dynamical methods for the null geodesics, and effective potential and shadows are studied in comparison with result of Kerr BH. The Sect.~\ref{S4} comprises the constraints on the spin of BH parameters in comparison with EHT data. In Sect.~\ref{S4a}, we investigate the impact of plasma on shadows. Lastly, in Sect.~\ref{S5}, we present a brief conclusion and future prospects. Note that, we consider $G=\hbar=c=1$ in our calculations, unless otherwise mentioned.

\section{The Static and Rotating Quantum Corrected Black Hole Metrics}\label{S2}
In this section, we review the basic concepts and the development of the quantum corrected Schwarzschild BH metric and its rotating counterpart. The Oppenheimer-Snyder model \cite{PhysRev.56.455} describes the collapse of dust matter. Since, the model describes the Big Bang singularity, therefore, it was proposed to consider a Big Bounce instead of the Big Bang \cite{doi:10.1177/002182869802900108}. It is also proposed that quantum gravity can resolve the singularity problem, therefore, a Big Bounce is also supported by LQC \cite{PhysRevLett.96.141301,YANG20091,PhysRevLett.121.081303}. A 4D spherically symmetric Ashtekar-Pawlowski-Singh (APS) metric \cite{PhysRevLett.96.141301} given as
\begin{equation}
\text{d}s^2_{\text{APS}}=-\text{d}\tau^2+a(\tau)^2\text{d}\tilde{r}^2+a(\tau)^2\tilde{r}^2\text{d}\Omega_2^2, \label{aps}
\end{equation}
where, $\text{d}\Omega_2^2=\text{d}\theta^2+\sin^2\theta\text{d}\phi^2$ is the metric of a 2-sphere and $a(\tau)$ satisfies a deformed Friedmann equation in terms of Hubble parameter $H$ given as
\begin{equation}
H^2:=\left(\frac{\dot{a}}{a}\right)^2=\frac{8\pi G\rho}{3}\left(1-\frac{\rho}{\rho_\text{c}}\right), \label{Friedmann}
\end{equation}
such that the energy density of the collapsing dust ball is given as $\rho=3M/\left(4\pi\tilde{r}_0^3a^3\right)$. The dot in Eq.~(\ref{Friedmann}) denotes the differentiation with respect to the proper time $\tau$. Moreover, the critical energy density $\rho_\text{c}$ causes the deformation with a constant value equal to $\sqrt{3}/\left(32\pi^2\gamma^3G^2\hbar\right)$. Note that $M$ is the mass of the dust ball with radius $a(\tau)\tilde{r}_0$, $\gamma$ is the Barbero-Immirzi parameter of LQG \cite{KrzysztofAMeissner_2004,MarcinDomagala_2004}, whereas, $G$ and $\hbar$ are Newton's and Planck's constants, respectively. The mass of the dust ball does not vary due to the conservation of the energy-momentum tensor. The classical regime corresponds to $\rho\ll\rho_\text{c}$, whereas, the energy density of the ball is never infinite for which the APS metric does not exhibit any singularity. Any particle inside the dust ball satisfies the inequality $0\leq\tilde{r}\leq\tilde{r}_0$.

The quantum Oppenheimer-Snyder (qOS) model is given by the metric
\begin{equation}
\text{d}s^2_{\text{qOS}}=-\left(1-F(r)\right)\text{d}t^2+\frac{\text{d}r^2}{1-H(r)}+r^2\text{d}\Omega_2^2. \label{qos}
\end{equation}
The interior region is described by the APS metric as a dust ball, and the exterior one is depicted by vacuum qOS metric. The $\theta$ and $\phi$ coordinates are same for both metrics, whereas, the coordinates $\tau$ and $\tilde{r}$ in the ball are matched on to the coordinates $t$ and $r$ in the exterior region. The dust interface $\tilde{r}=\tilde{r}_0$ is important region in APS spacetime, with coordinates $(t(\tau),r(\tau),\theta,\phi)$ in the other spacetime. Using the Israel junction conditions, which require continuity of the metrics and extrinsic curvature on the interface between dust and vacuum, the coordinates are matched as $(\tau,\tilde{r}_0,\theta,\phi)\sim(t(\tau),r(\tau),\theta,\phi)$, generating a quantum corrected Schwarzschild metric in LQG give by \cite{PhysRevLett.130.101501}
\begin{eqnarray}
\text{d}s^2_{\text{qOS}}&=&-\left(1-\frac{2M}{r}+\frac{\alpha M^2}{r^4}\right)\text{d}t^2\nonumber\\&&+\left(1-\frac{2M}{r}+\frac{\alpha M^2}{r^4}\right)^{-1}\text{d}r^2+r^2\text{d}\Omega_2^2, \label{lqg}
\end{eqnarray}
where, the quantum correction parameter $\alpha=16\sqrt{3}\pi\gamma^3l_\text{p}^2$ causes the deformation in the Schwarzschild metric, and $l_\text{p}=\sqrt{G\hbar}$ defines the Planck length. It is understood that the Barbero-Immirzi parameter has a fixed value $\gamma\approx0.2375$ \cite{KrzysztofAMeissner_2004,MarcinDomagala_2004} that gives a fixed value of the parameter $\alpha\approx1.1663$ under the assumption $\hbar=G=1=c$ in natural units.

At the interface, the radius of the dust ball being $a(\tau)\tilde{r}_0$ is equal to the radius $r(\tau)$ in the other spacetime, therefore, taking the derivative on both sides and then squaring gives the value of $H(r(\tau))$. Along the radial geodesic, when $\dot{r}=0$, we have $H(r)=0$ which gives the lower bound of the radial coordinate, that is,
\begin{eqnarray}
r_\text{b}=\left(\frac{\alpha M}{2}\right)^{\frac{1}{3}}. \label{lb}
\end{eqnarray}
This implies that the radius of the dust surface $a(\tau)\tilde{r}_0\in[r_\text{b},\infty)$ and thus $r\geq r_\text{b}$. The mass $M$ of the dust ball is now the mass of the quantum corrected BH that has a minimum value $M_{\text{min}}=\frac{16\gamma\sqrt{\pi\gamma}}{3\sqrt[4]{3}}$. Below this minimum value of $M$, there exist no horizon, however, two horizons exist for $M>M_{\text{min}}$ given as
\begin{eqnarray}
r_\pm=\frac{\zeta\left(1\pm\sqrt{2\zeta-1}\right)\sqrt{\alpha}}{\sqrt{\left(1+\zeta\right)\left(1-\zeta\right)^3}}, \label{hor}
\end{eqnarray}
where, $\zeta\in\left(1/2,1\right)$ is arbitrary parameter such that $4\alpha\zeta^4=M^2\left(1-\zeta^2\right)^3$.

Recently, in Refs.~\cite{refId0a,2024arXiv240916323L,2024arXiv241009198A,2024arXiv241011332V}, the authors have not considered the above-mentioned fixed value of the Barbero-Immirzi parameter, but accomplished their analyses by incorporating the variation of $\alpha$. It is now well understood that $\gamma$ has a fixed value, therefore, the minimum mass $M_{\text{min}}$ and the deformation parameter $\alpha$ also have fixed values. Therefore, only $M$ is the free parameter in the BH metric (\ref{lqg}). The extra term with $\alpha$ describes only the deviation in the quantum regime from the Schwarzschild metric depending on $M$. Generally, for each particular value of $M$, the deviation between the quantum corrected BH and the Schwarzschild BH is fixed due to the fact that $\alpha$ has a fixed value and cannot be varied around its prescribed value. Therefore, the studies mentioned above with variable $\alpha$ do not account for rigorous and feasible analyses. Since, $M>M_{\text{min}}\approx0.8314$ is a valid limit for the BH mass, therefore, we consider $M=1$ throughout this work to focus on influence of $a$ within the LQG regime.

The significance of studying rotating BHs can be highlighted through a comparison of their shadows with the EHT data on supermassive BHs as they are predominantly rotating in nature. Thus, considering rotating BHs allow us to establish a more accurate and rigorous analysis. The rotating counterpart of the metric (\ref{lqg}) with effective geometry was derived in Refs. \cite{2024arXiv240917099F,2024arXiv241009198A,2024arXiv241011332V} so that the quantum corrected Kerr BH in effective LQG in Boyer-Lindquist coordinates reads 
\begin{eqnarray}
\text{d}s_{\text{eff}}^2&=&-\frac{\Delta(r)-a^2\sin^2\theta}{\rho^2}\text{d}t^2+\frac{\rho^2}{\Delta(r)}\text{d}r^2+\rho^2\text{d}\theta^2\nonumber\\
&&+\frac{\left(r^2+a^2\right)^2\sin^2\theta-\Delta(r)a^2\sin^4\theta}{\rho^2}\text{d}\phi^2\nonumber\\
&&-\frac{2a\sin^2\theta\left(a^2+r^2-\Delta(r)\right)}{\rho^2}\text{d}t\text{d}\phi, \label{romet}
\end{eqnarray}
where,
\begin{eqnarray}
\Delta(r)&=&r^2+a^2-2Mr+\frac{\alpha M^2}{r^2}, \label{del}\\
\rho^2&=&r^2+a^2\cos^2\theta, \label{rho}
\end{eqnarray}
where, $a$ is the spin parameter. In the static case, the model does not admit classical covariance. However, it is proved that there is a modified form of covariance, since the constraint algebra closes, even if the structure functions are modified \cite{PhysRevD.102.106024}. Since, the covariance is modified in static case for the spherical symmetry, the rotating metric may not be exact line element for its static counterpart giving rise to an effective geometry. Moreover, it is not required at all that classical covariance needs to be preserved at the effective level. For this reason, the metric (\ref{romet}) serves as an effective metric and therefore describes a quantum corrected Kerr BH in effective LQG which encompasses the most probable properties of the exact rotating metric. For instance, the metric (\ref{romet}) reduces to the metric (\ref{lqg}) when $a=0$. By removing the quantum effects, the metric (\ref{romet}) reduces to the Kerr metric, just as the metric (\ref{lqg}) reduces to Schwarzschild one. Moreover, the metric function $\Delta(r)$ can be written as
\begin{eqnarray}
\Delta(r)=\Delta_{\text{Kerr}}(r)+\frac{\alpha M^2}{r^2}, \label{delK}
\end{eqnarray}
which shows that the metric (\ref{romet}) is a quantum corrected Kerr BH metric with an additional term comprising LQG parameter. We have mentioned that the parameter $\alpha$ has a fixed value and we consider $M=1$, thus, the only free parameter for the rotating metric is the spin $a$. Hence, we can rigorously analyze the deviation of quantum corrected Kerr BH from the Kerr BH. Like the Kerr metric, the quantum corrected Kerr metric also exhibits the time-translational and rotational invariance isometries corresponding to the Killing vector fields $\left(\partial_t\right)^\mu$ and $\left(\partial_\phi\right)^\mu$.

\begin{figure}[t!]
\centering
\subfigure{\includegraphics[width=0.43\textwidth]{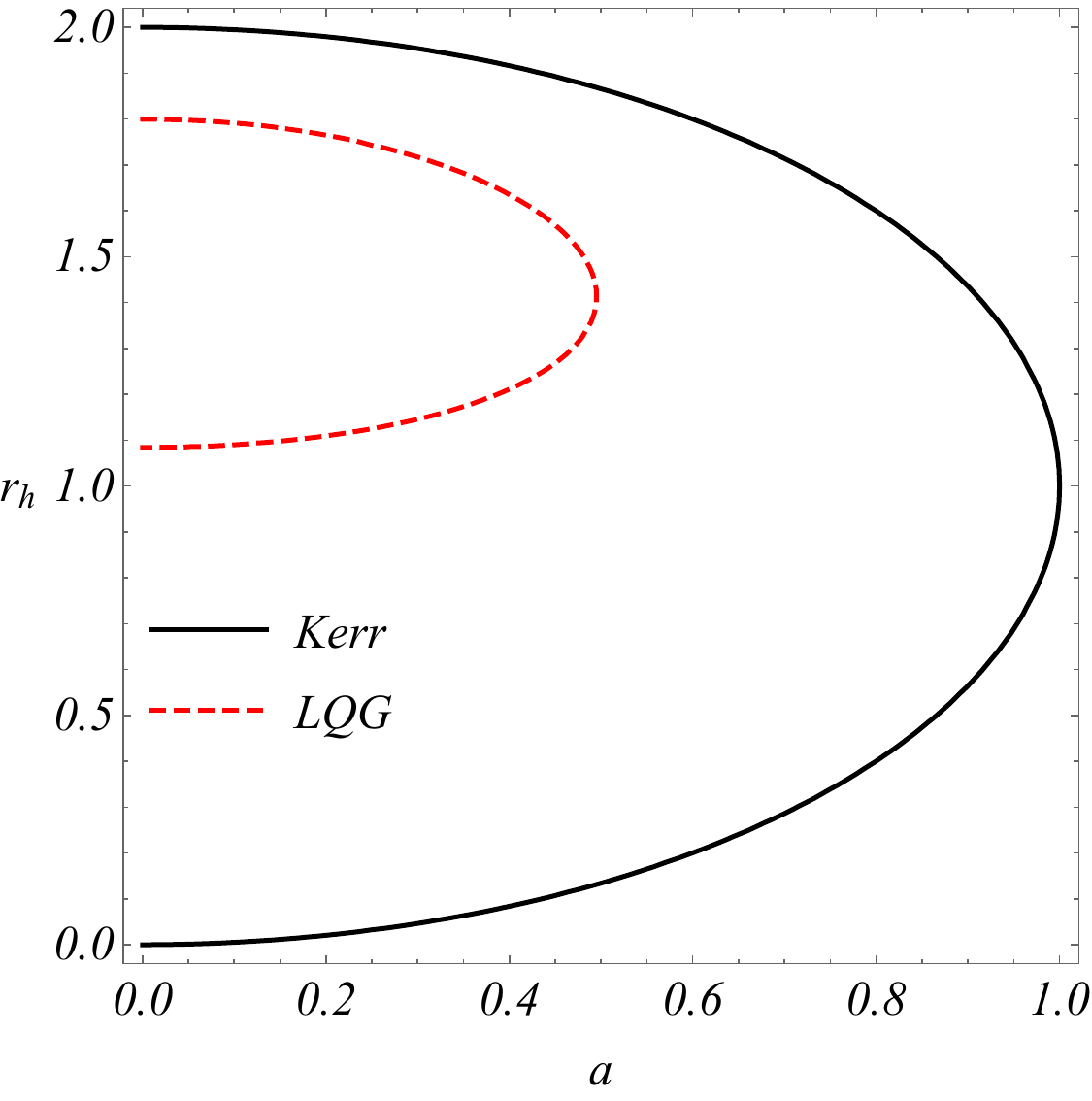}}
\caption{Horizon structure of Kerr and quantum corrected Kerr BHs with respect to $a$ for $M=1$ and $\alpha\approx1.1663$. \label{F1}}
\end{figure}

The last term in Eq.~(\ref{delK}) has no dependence on $a$, therefore, the horizon curve will be unique with a certain deviation from the horizon curve of Kerr BH. It is generated numerically by solving the equation $\Delta(r)=0$ for real and positive roots. A clear deviation in two curves is shown in Fig.~\ref{F1} in terms of horizon radii $r_\text{h}$ with respect to $a$. It can be seen that the event horizon of the quantum corrected Kerr BH is smaller than the event horizon of Kerr BH. While the Cauchy horizon of Kerr BH is significantly smaller than the event horizon of quantum corrected Kerr BH. Moreover, the extremal quantum corrected Kerr BH has spin $\sim$0.4952 which is marginally less than half of the extremal spin of Kerr BH.

\section{Unstable Null Orbits and Black Hole Shadow}\label{S3}
The photons emerging from a bright source may get trapped in unstable and stable circular null orbits in the vicinity of a BH. Some of them in the unstable orbits fall into the event horizon, while the rest scatter away to the infinity. This is how the optical image of the BH is formed, termed as shadow \cite{10.1093/mnras/131.3.463,1983mtbh.book.....C}. These orbits are characterized by an effective potential function. Therefore, we derive the null geodesic equations to study the effective potential and shadows of quantum corrected Kerr BH in effective LQG, in order to understand the influence of LQG via Barbero-Immirzi parameter in terms of unstable orbits and optical images. For the quantum corrected Kerr BH in effective LQG, the null geodesic equations can be obtained by employing the Hamilton-Jacobi formalism \cite{PhysRev.174.1559} which has been widely incorporated in the literature over the years. One can also begin with the Lagrangian or Hamiltonian methods, generating two constants of motion, the energy $E$ and the angular momentum $L$ along with mass of the particle under motion. The geodesic equations become completely integrable if we introduce a fourth constant of motion. The separation of variables arising in Hamilton-Jacobi formalism enables us to introduce a constant known as the Carter constant \cite{PhysRev.174.1559}. We consider the Hamilton-Jacobi equation
\begin{equation}
2\partial_\tau S=-g^{\mu\nu}\partial_{x^\mu}S\partial_{x^\nu}S \label{hj}
\end{equation}
with the Jacobi action of the form
\begin{equation}
S=\frac{1}{2}m_\text{p}^2\tau-Et+L\phi+A_r(r)+A_\theta(\theta), \label{jact}
\end{equation}
where, $\tau$ being the proper time is considered as an affine parameter and $m_\text{p}=0$ is the photon mass. The functions $A_r(r)$ and $A_\theta(\theta)$ are arbitrary functions to be determined. The constants $E=-p_t$ and $L=p_\phi$ can be obtained from the relation $p_\mu=g_{\mu\nu}\dot{x}^\nu$. Ultimately, the geodesic equations come out to be of the usual form as for the Kerr metric given as \cite{1983mtbh.book.....C}
\begin{align}
\rho^2\dot{t}&=\frac{r^2+a^2}{\Delta(r)}\left(E\left(r^2+a^2\right)-aL\right)+a\left(L-aE\sin^2\theta\right), \label{teq}\\
\rho^2\dot{r}&=\pm\sqrt{\mathcal{R}(r)}, \label{req}\\
\rho^2\dot{\theta}&=\pm\sqrt{\Theta(\theta)}, \label{theq}\\
\rho^2\dot{\phi}&=\frac{a}{\Delta(r)}\left(E\left(r^2+a^2\right)-aL\right)+\left(L\csc^2\theta-aE\right), \label{pheq}
\end{align}
where,
\begin{align}
\mathcal{R}(r)&=\left(\left(r^2+a^2\right)E-aL\right)^2-\Delta(r)\left(\mathcal{Z}+(L-aE)^2\right), \label{Rfun}\\
\Theta(\theta)&=\mathcal{Z}+\cos^2\theta\left(a^2E^2-L^2\csc^2\theta\right), \label{thfun}
\end{align}
where, $\mathcal{Z}$ denotes the Carter constant. The function $\mathcal{R}(r)$ is of prime importance in studying the effective potential and behavior of unstable orbits as it connects the radial geodesic equation with the effective potential. In this case, it can also be expressed as
\begin{eqnarray}
\mathcal{R}(r)=\mathcal{R}_{\text{Kerr}}(r)-\frac{\alpha\left(\mathcal{Z}+(L-aE)^2\right)M^2}{r^2}, \label{RK}
\end{eqnarray}
where, the second term on right hand side of Eq.~(\ref{RK}) is the deviation factor of $\mathcal{R}(r)$ from $\mathcal{R}_{\text{Kerr}}(r)$. The deviation in the metric function is fixed as it is independent of free parameters, however, the deviation in $\mathcal{R}(r)$ is dependent on $a$ due to which it varies for each case of BH spin. Note that the quantum correction does not affect the function $\Theta(\theta)$. The photon moving in a circular orbit is subjected to centripetal force to keep it in its orbit with an opposing force called the centrifugal force. The centrifugal force corresponds to a potential known as centrifugal potential that together with real potential makes up the effective potential. From the radial null geodesic equation, it can be written in terms of effective potential for Kerr BH for equatorial trajectories as
\begin{eqnarray}
V^{\text{eff}}(r)=V^{\text{eff}}_{\text{Kerr}}(r)+\frac{\alpha\left(\mathcal{Z}+(L-aE)^2\right)M^2}{2r^6}. \label{Veff}
\end{eqnarray}
As in the case of $\mathcal{R}(r)$, the quantum correction term in the effective potential also depends on spin $a$ that causes a variable deformation in the null orbits. This deviation is generally in terms of either shrinking or expansion of null orbits. However, depending on the type of spacetime metric, the shape of deformation term in metric function and effective potential functions, we can determine whether the null orbits shrink or expand. For a concrete and robust result under certain assumptions, we establish a theorem and present a simple proof to it as follows:

\textit{Theorem: Suppose a Kerr-like metric described by $\Delta(r)=\Delta_{\text{Kerr}}(r)+b_1r^p$ with $V^{\text{eff}}(r)=V_{\text{Kerr}}^{\text{eff}}(r)+b_2r^q$ such that $r^p$ and $r^q$ be decreasing functions in an open interval $\mathcal{I}=(0,s)\subseteq\mathbb{R}^+$, and $b_1,b_2\in\mathbb{R}^+$. If an unstable null orbit for Kerr BH exists at a radial distance $r_k\in\mathcal{I}$, then the unstable null orbit corresponding to $\Delta(r)$ exists at some $r_0\in\mathcal{I}$ such that $r_0<r_k$.}
\begin{figure}[t!]
\centering
\subfigure{\includegraphics[width=0.43\textwidth]{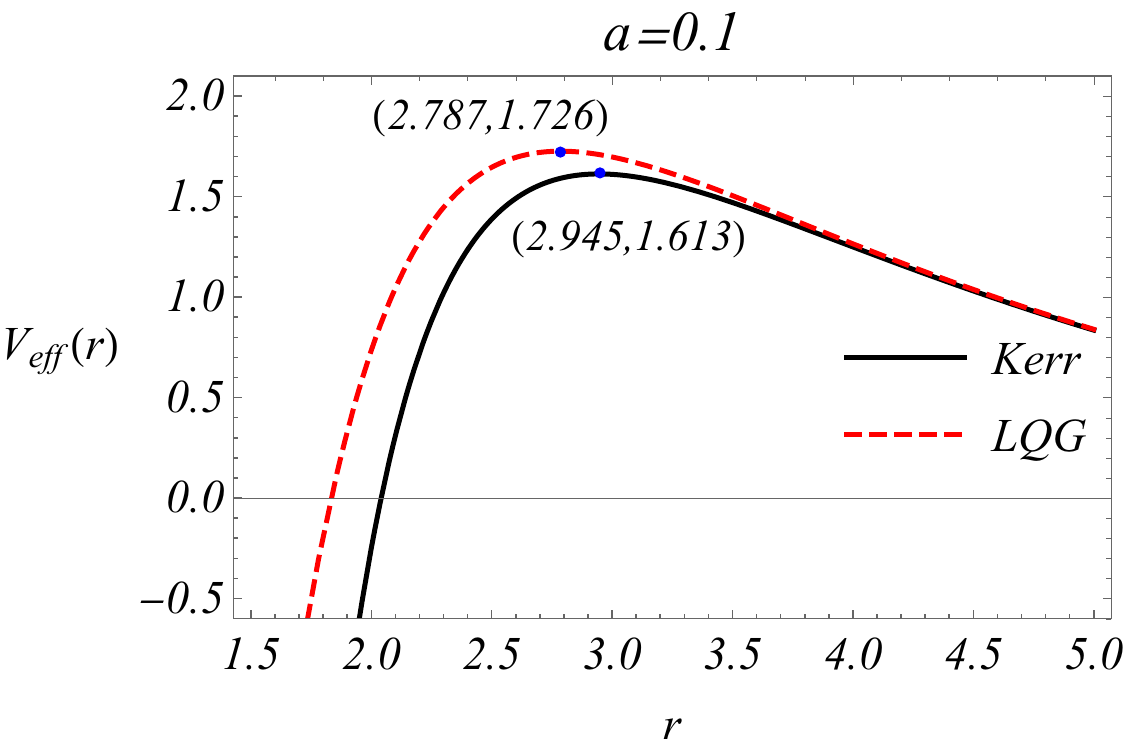}}
\subfigure{\includegraphics[width=0.43\textwidth]{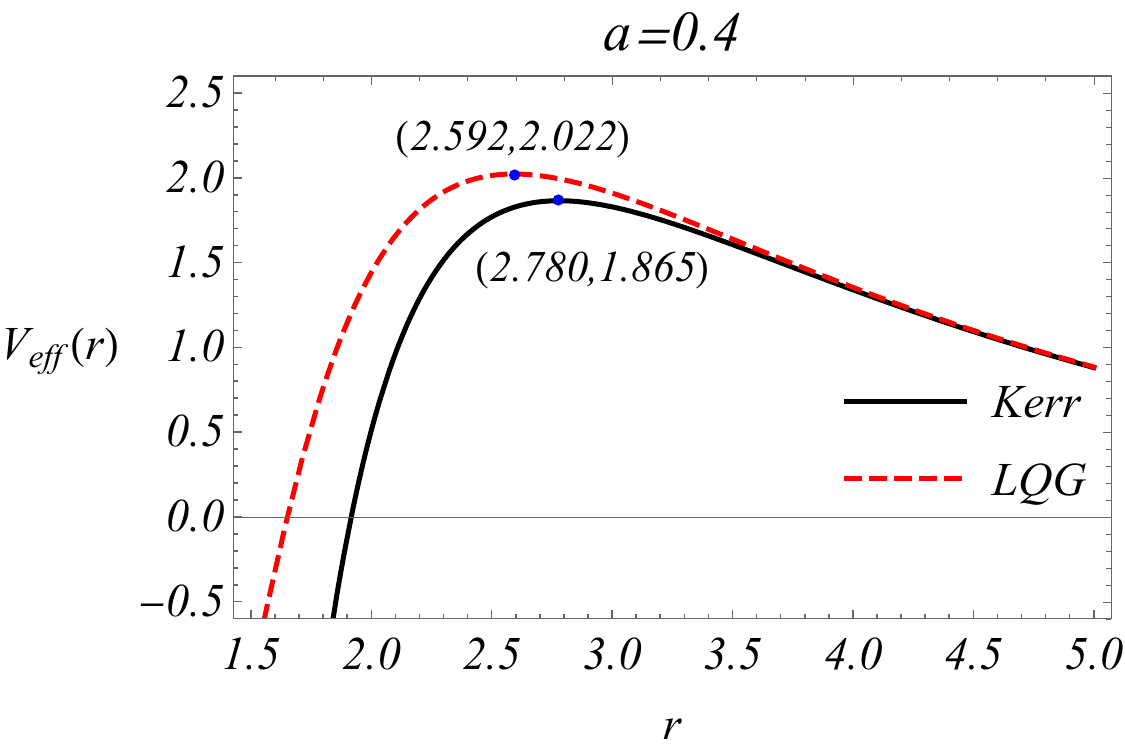}}
\subfigure{\includegraphics[width=0.43\textwidth]{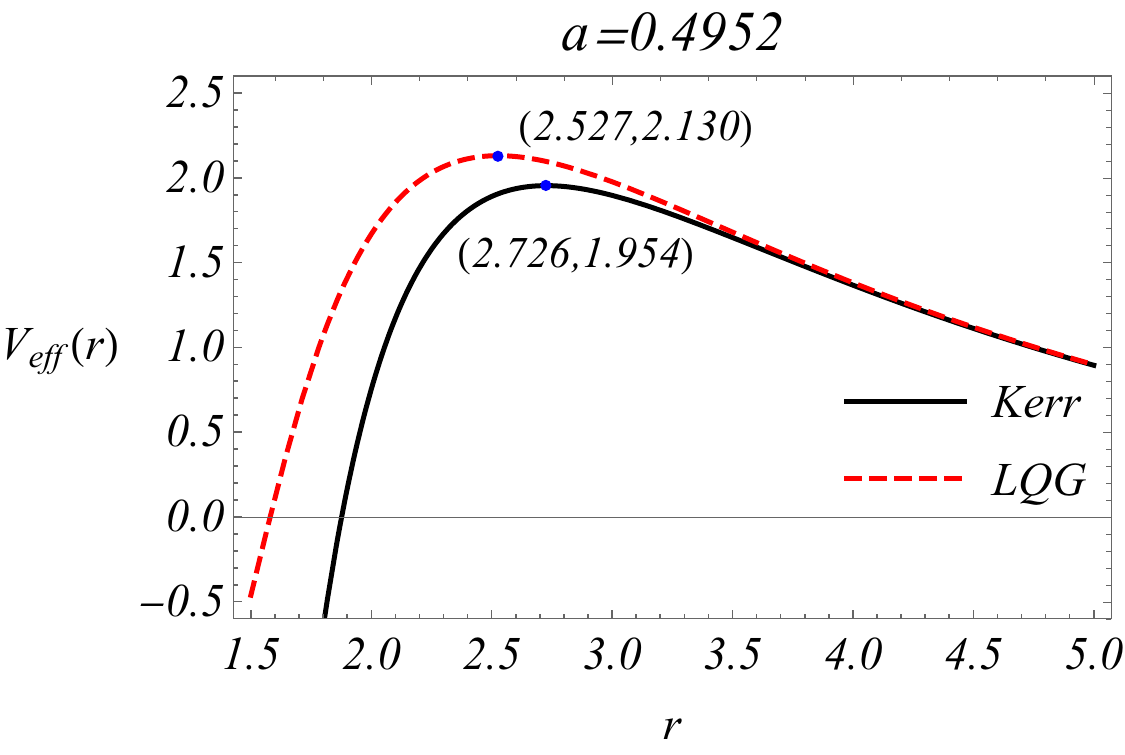}}
\caption{Behavior of effective potential and unstable null orbits for Kerr and quantum corrected Kerr BHs for different values of $a$ with $M=1$ and $\alpha\approx1.1663$. \label{F2}}
\end{figure}
\textit{Proof:} Given that $\Delta(r)=\Delta_{\text{Kerr}}(r)+b_1r^p$ and $V^{\text{eff}}(r)=V_{\text{Kerr}}^{\text{eff}}(r)+b_2r^q$ with decreasing functions $r^p$ and $r^q$ such that $p,q\in\mathbb{R}^-$, and $b_1,b_2\in\mathbb{R}^+$ depend only on spin $a$. Now, if an unstable null orbit for Kerr BH exists at some $r_k\in\mathcal{I}$, then the function $V_{\text{Kerr}}^{\text{eff}}(r)$ is concave at $r_k$ that is $\partial_r^2V_{\text{Kerr}}^{\text{eff}}(r_k)<0$ and $\partial_rV_{\text{Kerr}}^{\text{eff}}(r_k)=0$. Since, $r^q$ is decreasing $\forall~r\in\mathcal{I}$, therefore, $\partial_rV^{\text{eff}}(r_k)<0$. From this, one of the following two possibilities holds:
\begin{itemize}
\item[\textit{i})] If $V^{\text{eff}}(r)$ is concave at $r_0$, then $\exists~r_0\in\mathcal{I}$ such that $r_0<r_k$ and $\partial_rV^{\text{eff}}(r_0)=0$.
\item[\textit{ii})] If $V^{\text{eff}}(r)$ is convex at $r_0$, then $\exists~r_0\in\mathcal{I}$ such that $r_0>r_k$ and $\partial_rV^{\text{eff}}(r_0)=0$.
\end{itemize}
To prove the theorem, we either prove $(i)$ holds or $(ii)$ does not hold. Contrarily suppose that $(ii)$ holds, then $\partial_rV_{\text{Kerr}}^{\text{eff}}(r_0)>0$ but $V_{\text{Kerr}}^{\text{eff}}(r)$ must be decreasing at $r_0$ under the given statement. This gives a contradiction which implies that $(ii)$ does not hold and thus $(i)$ holds automatically. Therefore, it proves that the unstable null orbits shrink under the given assumptions. Note that we have not proved $r_0\in\mathcal{I}$. Since, the lower bound of the interval $\mathcal{I}$ is $0$ and $r_0<r_k$, therefore, $r_0\in\mathcal{I}$. We have also omitted the case when $r_0=r_k$ because $b_1,b_2\neq0$. Moreover, the interval $\mathcal{I}$ is arbitrary and may have multiple extrema, however, the local maximum corresponding to the unstable orbit is unique.

The result proved in the above theorem can be viewed in Fig.~\ref{F2} in terms of location of peaks of the curves. That is, the unstable null orbits for quantum corrected Kerr BH are smaller than the unstable null orbits for Kerr BH for all cases. Moreover, for each case of extremal spin of quantum corrected Kerr BH, the difference between the sizes of unstable orbits for both BHs increases, which is specifically due to the presence of spin parameter in the deviation term in Eq.~(\ref{Veff}). Furthermore, with increase in spin, the unstable orbits of quantum corrected and Kerr BHs shrink. 

The effective potential function governs the behavior of timelike and null orbits around a BH. Whereas, the types of these orbits, whether stable or unstable, are determined mathematically by solving the equations $V^{\text{eff}}(r_\text{p})=0=\partial_rV^{\text{eff}}(r_\text{p})$ and then identifying the concavity or convexity of the function as $\partial_r^2V^{\text{eff}}(r_\text{p})<0$ for unstable orbits or $\partial_r^2V^{\text{eff}}(r_\text{p})>0$ for stable orbits. Since, the photons are trapped in circular orbits and these orbits together in all orientations make up a sphere whose radius is denoted by $r_\text{p}$. On the surface of sphere, the radial component of these photons satisfies the equation $r=constant$. Therefore, we have $\dot{r}=0=\ddot{r}$ which is equivalent to $\mathcal{R}(r_\text{p})=0=\partial_r\mathcal{R}(r_\text{p})$ by Eq.~(\ref{req}). Hence, solving the equations for critical orbits in terms of effective potential is equivalent to solving in terms of $\mathcal{R}(r)$. These equations give the values of impact parameters $\xi=L/E$ and $\eta=\mathcal{Z}/E^2$ as a function of arbitrary $r_\text{p}$ given as
\begin{align}
\xi(r_\text{p})&=\xi_{\text{Kerr}}(r_\text{p})+\frac{2\alpha M^2r_\text{p}\left(\Delta_{\text{Kerr}}(r_\text{p})+r_\text{p}\Gamma\right)}{a\Gamma\left(\alpha M^2-r_\text{p}^3\Gamma\right)}, \label{xi}\\
\eta(r_\text{p})&=\eta_{\text{Kerr}}(r_\text{p})+\frac{4\alpha M^2r_\text{p}^3}{\Gamma^2\left(\alpha M^2-\Gamma r_\text{p}^3\right)^2}\Bigg[2\Gamma r_\text{p}^3(M-\Gamma)\nonumber\\&-\alpha M^3+\frac{r_\text{p}^2(M-2\Gamma)\left[M\left(\alpha M-3r_\text{p}^2\Gamma\right)+\Gamma r_\text{p}^3\right]}{a^2}\Bigg], \label{eta}
\end{align}
where, $\Gamma=r_\text{p}-M$. For $\alpha=0$, the quantum correction terms in Eqs.~(\ref{xi}) and (\ref{eta}) vanish, and the impact parameters for Kerr BH are recovered. These impact parameters then determine the celestial coordinates \cite{2009PhRvD..80b4042H}
\begin{align}
X(r_\text{p})=-\lim_{\substack{r\to\infty\\\theta\to\theta_0}}r^2\sin\theta\frac{d\phi}{dr}, \quad Y(r_\text{p})=\lim_{\substack{r\to\infty\\\theta\to\theta_0}}r^2\frac{d\theta}{dr} \label{cel}
\end{align}
to sketch a projective 2D image of the BH shadow, given in the form
\begin{align}
X(r_\text{p})&=-\xi(r_\text{p})\csc\theta_0, \label{celx}\\
Y(r_\text{p})&=\pm\sqrt{\eta(r_\text{p})+a^2\cos^2\theta_0-\xi^2(r_\text{p})\cot^2\theta_0}. \label{cely}
\end{align}
The limits in Eq.~(\ref{cel}) correspond to the radial and angular components of observer's location. For an equatorial observer, the celestial coordinates reduce to
\begin{equation}
X(r_\text{p})=-\xi(r_\text{p}), \quad Y(r_\text{p})=\pm\sqrt{\eta(r_\text{p})}. \label{cele}
\end{equation}
Since, the shadows are formed as circular contours and often deformed for the rotating BH cases. Therefore, one can assume the condition $Y(r_\text{p})=0$ corresponding to the extreme points on the shadow contour on the $X$-axis of the celestial plane which ultimately correspond to the extreme values of the interval $[r^{\text{min}}_\text{p},r^{\text{max}}_\text{p}]$ for the photon sphere. However, the photon sphere around a static BH has a unique width which cannot be considered as a parameter. To deal with this, one can modify the impact parameter $\eta$ in the form $\eta_\text{m}=\left(\mathcal{Z}+(L-aE)^2\right)/E^2$ and consider $\xi$ as the parameter. The extreme values of the interval containing $\xi$ are determined by solving the equation $\Theta(\theta_0)=0$. The celestial coordinates in Eq.~(\ref{cele}) can also be expressed as
\begin{figure}[t!]
\centering
\subfigure{\includegraphics[width=0.36\textwidth]{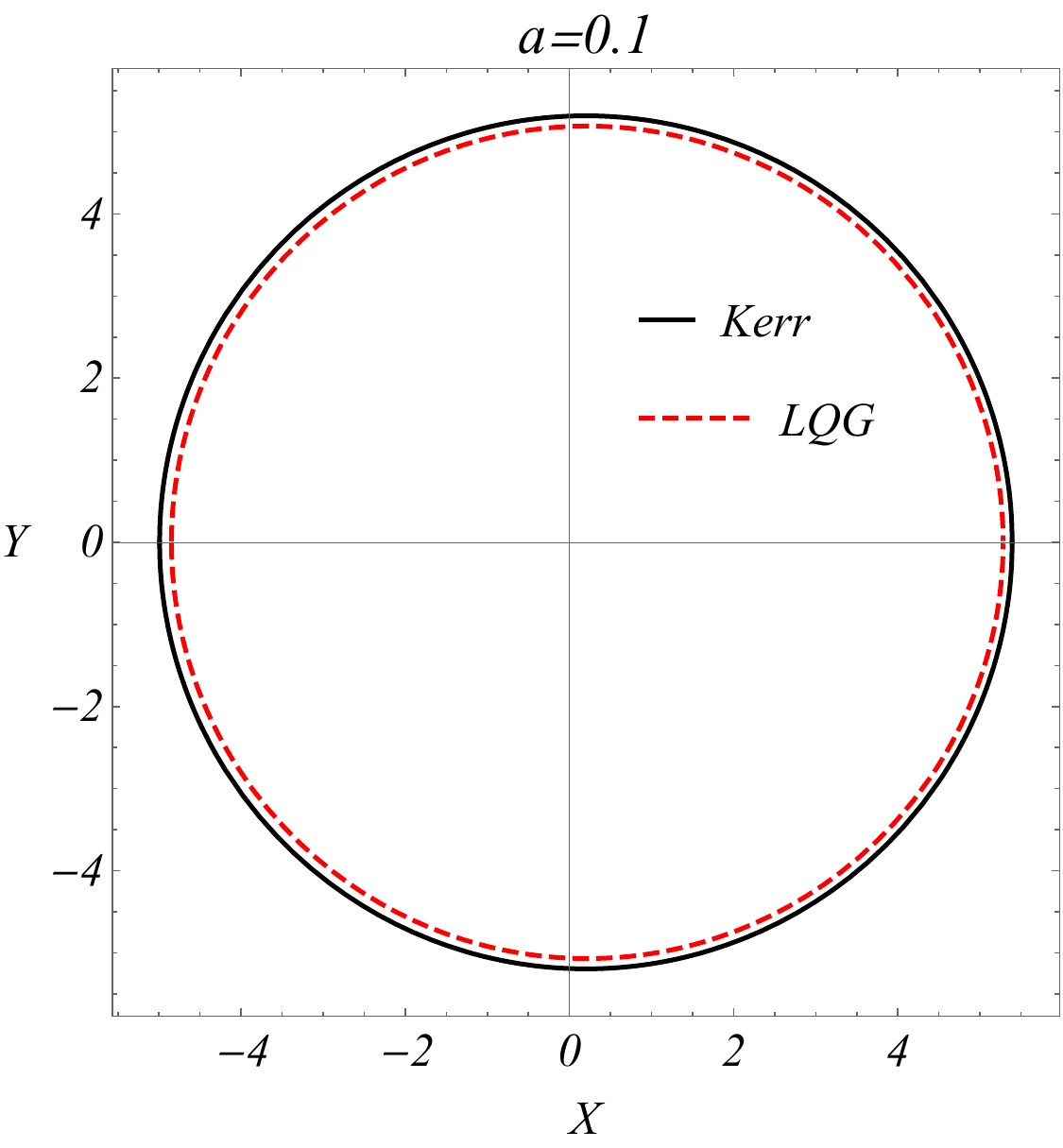}}
\subfigure{\includegraphics[width=0.36\textwidth]{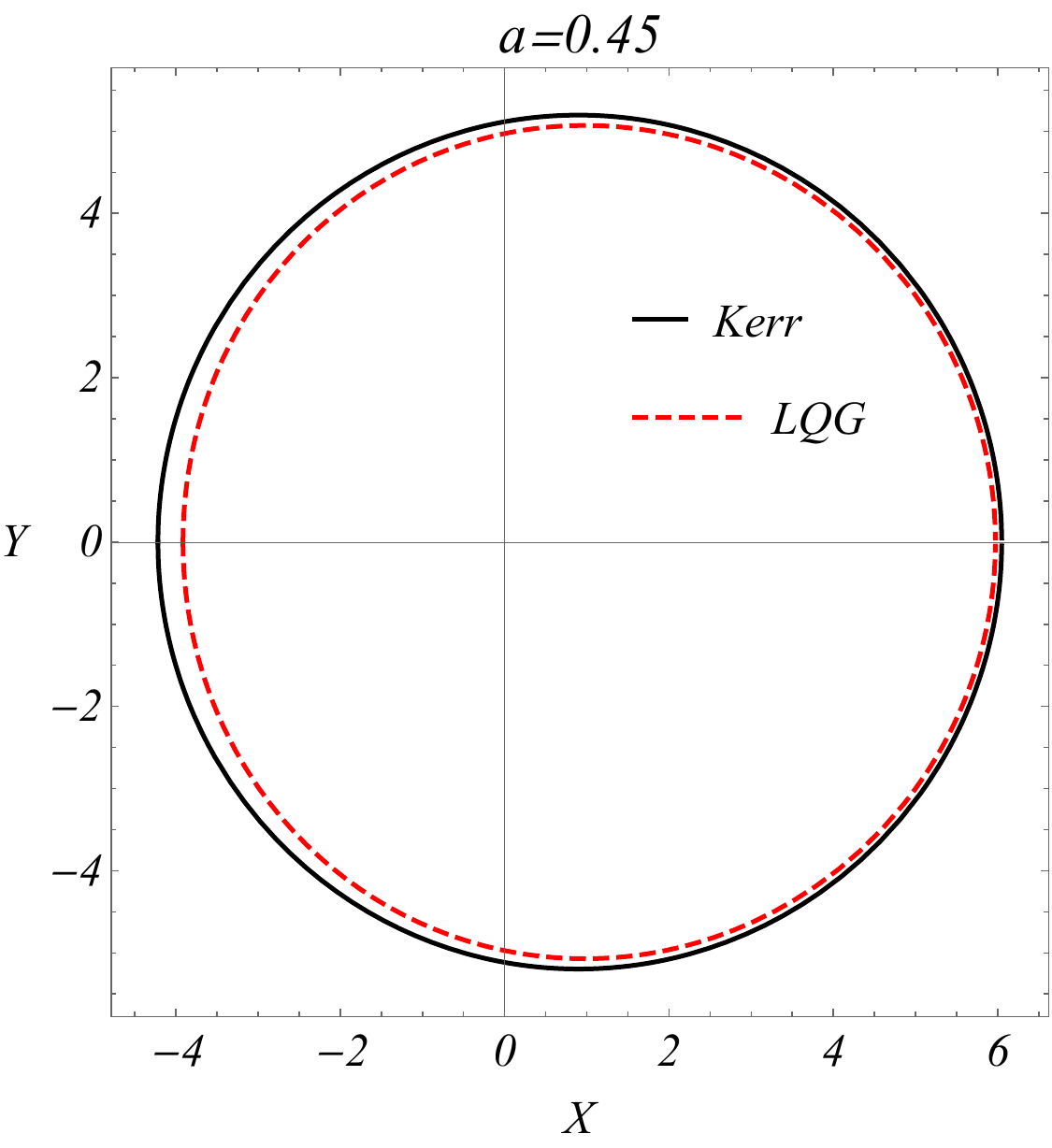}}
\subfigure{\includegraphics[width=0.36\textwidth]{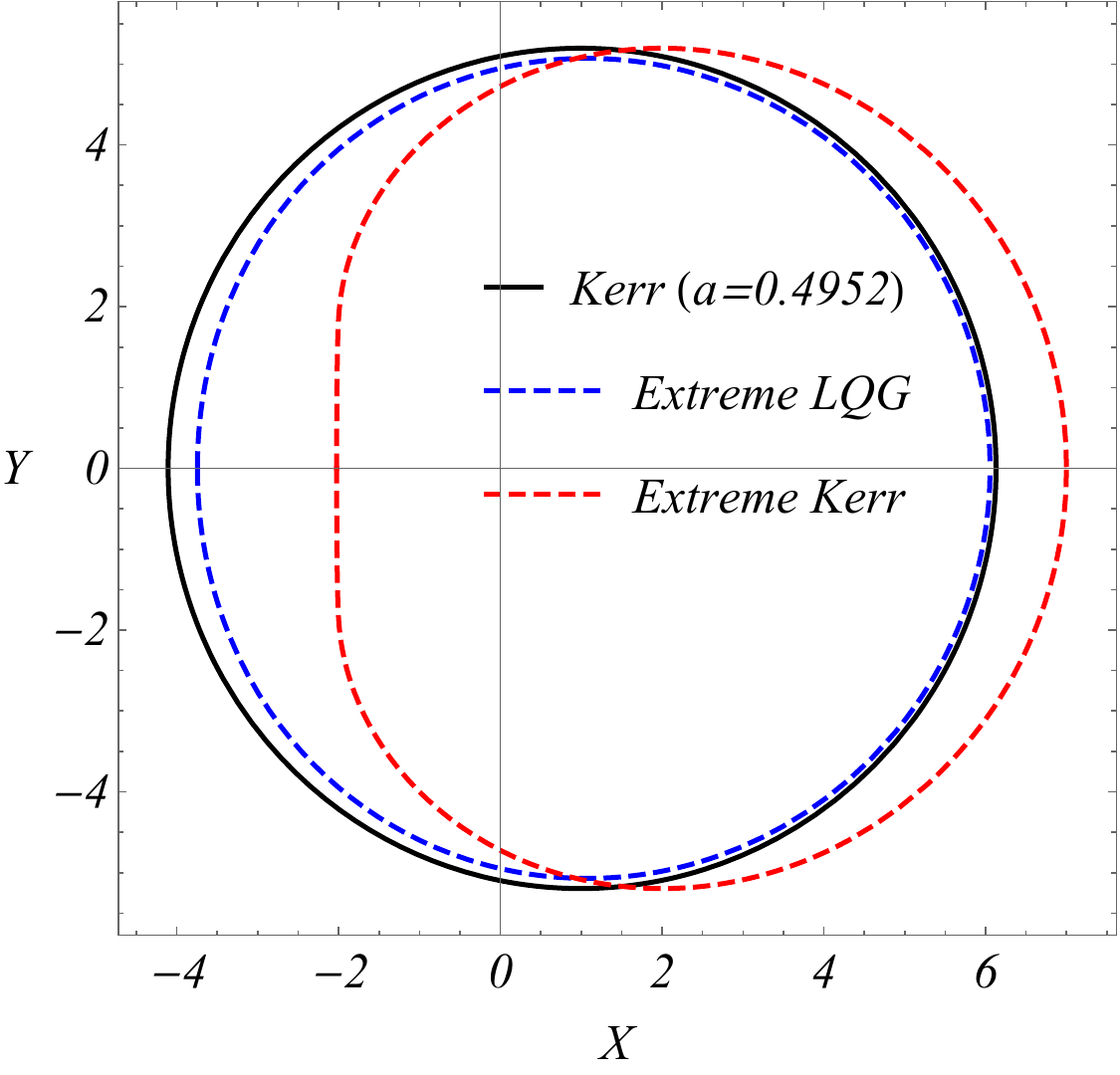}}
\caption{Behavior of shadows for Kerr and quantum corrected Kerr BHs for different values of $a$ with $M=1$ and $\alpha\approx1.1663$, visualized by an equatorial observer at radial infinity. \label{F3}}
\end{figure}
\begin{align}
X(r_\text{p})&=X_{\text{Kerr}}(r_\text{p})-\frac{2\alpha M^2r_\text{p}\left(\Delta_{\text{Kerr}}(r_\text{p})+r_\text{p}\Gamma\right)}{a\Gamma\left(\alpha M^2-r_\text{p}^3\Gamma\right)}, \label{xk}\\
Y(r_\text{p})&=Y_{\text{Kerr}}(r_\text{p})\pm\frac{2\alpha M^2r_\text{p}^3}{\Gamma^2\sqrt{\eta_{\text{Kerr}}(r_\text{p})}\left(\alpha M^2-\Gamma r_\text{p}^3\right)^2}\nonumber\\&\Bigg[\frac{r_\text{p}^2(M-2\Gamma)\left[M\left(\alpha M-3r_\text{p}^2\Gamma\right)+\Gamma r_\text{p}^3\right]}{a^2}\nonumber\\&+2\Gamma r_\text{p}^3(M-\Gamma)-\alpha M^3\Bigg]\mp\text{higher order terms}, \label{yk}
\end{align}
which further reduce to the celestial coordinates for Kerr BH under the limit $\alpha=0$. The extra terms on the right hand side of Eqs.~(\ref{xk}) and (\ref{yk}) describe the deviation of shadow due to LQG effects from the shadow of Kerr BH. The $Y$-coordinate in Eq.~(\ref{yk}) is expressed in terms of an alternating infinite series contributing to the deviation of the shadow from the shadow of Kerr BH. To quantify this deviation in the shadow, we plot some shadow contours for a few cases of spin parameter values as given in Fig.~\ref{F3}. From the difference of sizes of the Kerr and quantum corrected Kerr BHs for all cases, one can deduce that the quantum corrected Kerr BH will appear smaller to a visualizing observer at infinity as compared to the Kerr BH. As the value of spin parameter increases, the distortion starts appearing in the shadows of quantum corrected Kerr BH. That is, the difference between the shadow contours on the left side increases with increase in spin. It suggests that the spin parameter influences the elongation of shadows for quantum corrected Kerr BH more than the Kerr BH up to $a=0.4952$. The third plot with three contours show a huge difference between the shadows of quantum corrected and Kerr BHs for their respective extreme spin values. Therefore, one may not expect to visualize a perfectly flattened shadow for extreme quantum corrected Kerr BH. Moreover, the deviation in shadow contours for $a=0.4952$ for both BHs can also be measured. For smaller values of spin, both BHs are centered closer to the origin and will certainly be centered at origin for $a=0$.
\begin{figure}[t!]
\centering
\subfigure{\includegraphics[width=0.41\textwidth]{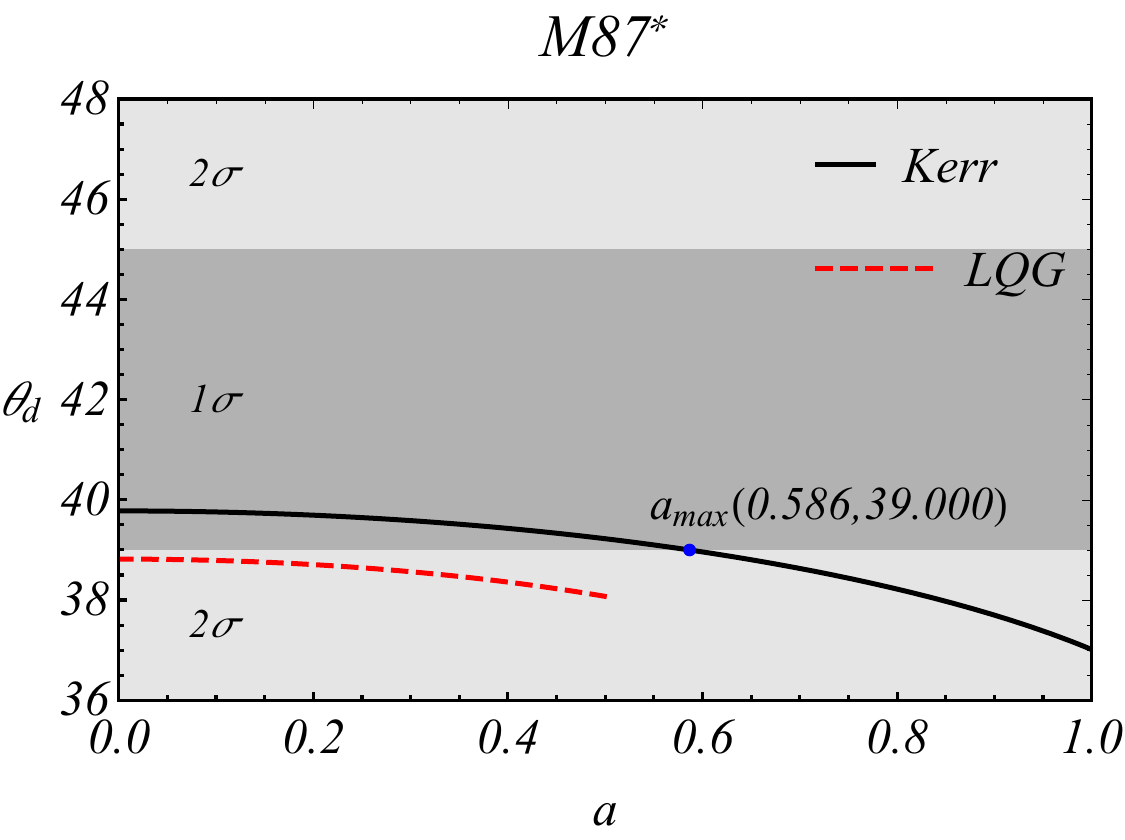}}
\subfigure{\includegraphics[width=0.41\textwidth]{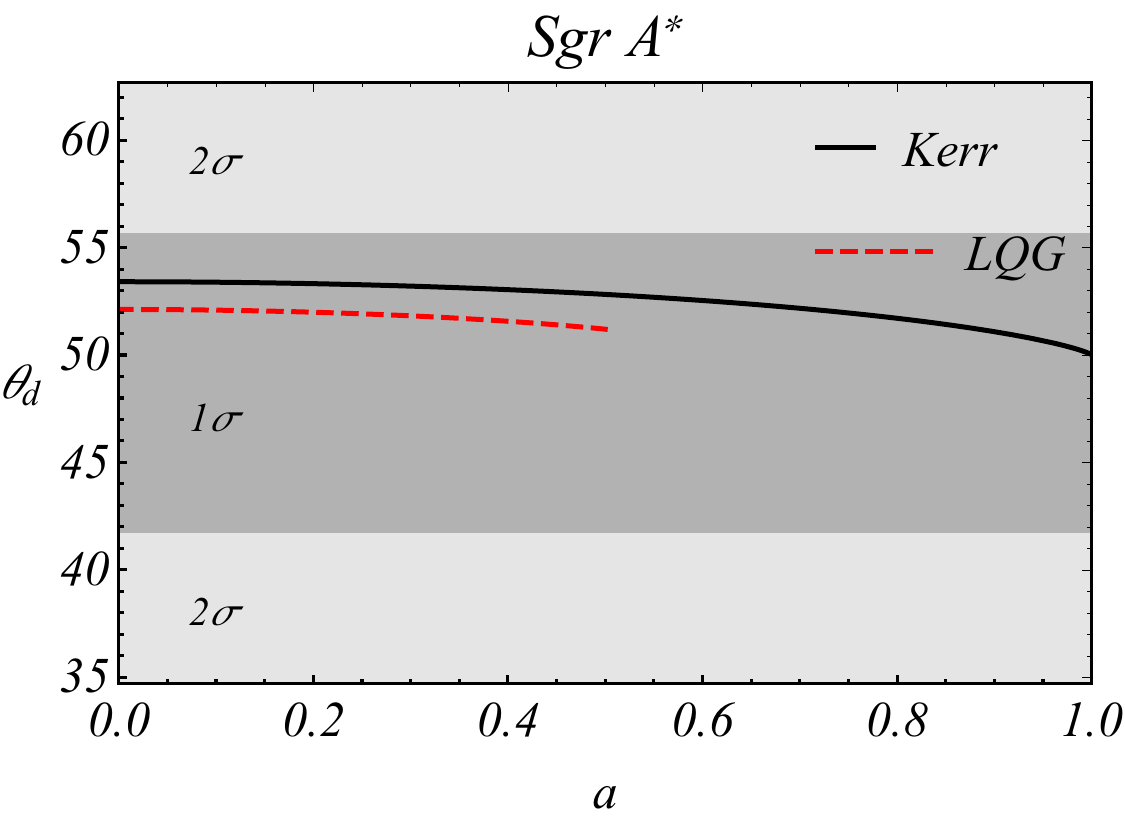}}
\caption{Comparison of shadow angular diameter $\theta_\text{d}$ for quantum corrected Kerr BH (dashed red curve) and Kerr BH (solid black curves) with the EHT data for M87* (at inclination angle of $17^\circ$) and Sgr A* (at inclination angle of $45^\circ$) for the bounds on spin $a$ within 1-$\sigma$ intervals. \label{F4}}
\end{figure}

\section{Comparison With EHT Data}\label{S4}
In this section, we will explore the constraints on the spin parameter $a$ for both Kerr and quantum corrected Kerr BHs, using observational data from the EHT collaborations. These observations, focusing on M87* and Sgr A*, will allow us to determine the limits on the spin values of these BHs. We will compare the constraints obtained for both BHs to gain insights into how LQG might influence the characteristics of the rotating BH, providing a deeper understanding of the potential effects of quantum gravity on BHs. To do this, we calculate the angular radii of the shadows of both BHs to establish a comparative analysis with the angular radii of M87* and Sgr A*. Corresponding to such a bound on the spin parameter, the BH is considered to mimic either M87* or Sgr A* if the angular diameter of BH shadow falls within 1-$\sigma$ interval. This study specifically examines rotating BHs, as supermassive BHs are naturally expected to exhibit significant rotational features due to their formation and evolution processes. A coordinate-independent formalism, commonly referred to as the Kumar-Ghosh method \cite{10.1093/mnras/stv2079,Kumar_2020a}, is employed, where the shadow area is utilized which is defined as
\begin{eqnarray}
A_{\text{sh}}=2\int_{r_-}^{r_+}dx^rY(r)\partial_rX(r). \label{area}
\end{eqnarray}
The values $r_+$ and $r_-$ represent the size of the retrograde and prograde stable circular orbits as measured from the origin, respectively. Suppose that the separation between the BH and observer is described by the linear distance $d$, then the diameter of the BH shadow is measured as \cite{Kumar_2020,Afrin_2023}
\begin{eqnarray}
\theta_\text{d}=\frac{2}{d}\sqrt{\frac{A_{\text{sh}}}{\pi}}. \label{ang}
\end{eqnarray}
By utilizing the relations (\ref{area}) and (\ref{ang}), the angular diameter of the BH shadow can be written in terms of spin $a$, $r_\text{p}$ and $\theta_0$. For the comparison of the shadows, the EHT data determines the distance $d$ of Earth from M87* and Sgr A*, the mass $M$ and the shadow size $\theta_\text{d}$ of M87* and Sgr A*. For M87*, we get $d=16.8\text{Mpc}$, $M=6.5\times10^9\text{M}_\odot$ and $\theta_\text{d}=42\pm3\mu\text{as}$ \cite{EHT2019,Akiyama_2019,Akiyama_2019a}. Whereas, for Sgr A*, we get $d=8\text{kpc}$, $M=4\times10^6\text{M}_\odot$ and $\theta_\text{d}=48.7\pm7\mu\text{as}$ \cite{Ehtl12_2022,Eht17_2022}. Here, $\text{M}_\odot$ denotes the solar mass, kpc and Mpc stand for kilo and mega parsec, and $\mu\text{as}$ stands for micro arcsec. For simplicity, we have not considered uncertainties in the measurements of mass and distance. The EHT conducted the observations at the inclination angles of $17^\circ$ for M87* and $<50^\circ$ for Sgr A*. Therefore, we will also consider the angles of $17^\circ$ for M87* and $45^\circ$ for Sgr A* for the calculations of shadows.

The impact of LQG that deviates the quantum corrected Kerr BH from Kerr BH is obvious in Fig.~\ref{F4}. We have plotted the shadow angular diameters of the Kerr and quantum corrected Kerr BHs with respect to spin $a$ and compared it with M87* and Sgr A*. The 1-$\sigma$ uncertainty levels are indicated by dark gray regions. For the comparison with M87*, we found that the shadow diameter of Kerr BH lies within 1-$\sigma$ uncertainty level for $0\leq a\lesssim0.586$, while, for other values of $a$, the shadow diameter of Kerr BH lies within 2-$\sigma$ uncertainty level. The upper bound of spin is denoted by $a_\text{max}\approx0.586$ at which the transition of shadow diameter is observed from 1-$\sigma$ level to 2-$\sigma$. Therefore, the Kerr BH can be regarded identical with M87* for $0\leq a\lesssim0.586$. However, the quantum corrected Kerr BH does not mimic M87* as its shadow diameter lies within 2-$\sigma$ uncertainty level for all values of spin which is a tremendous impact of LQG on quantum corrected Kerr BH. Generally, the 2-$\sigma$ uncertainty level is also considered for the comparison, however, we ignore this region in order to reduce the possibility of uncertainty. This essentially eliminates the possibility for the quantum corrected Kerr BH to behave like M87*. However, an entirely opposite impact of LQG is visualized for the case of Sgr A*. The Kerr BH behaves identical with Sgr A* for all values of spin as its shadow diameter lies within 1-$\sigma$ uncertainty level and $\theta_\text{d}$ approaches the median value $48.7\mu\text{as}$ as $a$ approaches its maximal value. For the quantum corrected Kerr BH, the shadow diameter also lies within 1-$\sigma$ uncertainty level. However, under the influence of LQG, the shadow diameter for quantum corrected Kerr BH is closer to the median value $48.7\mu\text{as}$ for each value of spin. Therefore, one may regard the quantum corrected Kerr BH to mimic Sgr A* for all values of spin which is more likely than the Kerr BH within its interval of spin.

\section{Effect of Plasma on Shadow}\label{S4a}
We know that astrophysical BHs are generally surrounded by plasma. Moreover, excited states of matter at high temperatures with their constituent particles (positive and negative charges) may have significant impact on the quantum corrected Kerr BH at Planck scales. This being an important aspect in theoretical physics, we investigate the influence of plasma on the appearance of quantum corrected Kerr BH. We consider pressureless and non-magnetized plasma with distribution functions dependent on $r$ and $\theta$. We begin by considering the plasma electron frequency defined as
\begin{equation}
\omega_\text{p}(r,\theta)=\frac{4\pi e^2}{m_\text{e}}N_\text{e}(r,\theta) \label{13p}
\end{equation}
that modifies the Hamiltonian describing the photon motion in a plasma medium given as \cite{PhysRevD.95.104003}
\begin{equation}
\mathcal{H}=\frac{1}{2}\left[g^{\mu\nu}p_\mu p_\nu+\omega^2_\text{p}(r,\theta)\right], \label{11p}
\end{equation}
where $e$, $m_\text{e}$ and $N_\text{e}$ are the charge, mass and number density of the electron, respectively. The plasma electron frequency is related to the refractive index $n(r,\theta)$ as
\begin{equation}
n(r,\theta)=\sqrt{1-\frac{\omega^2_\text{p}(r,\theta)}{\omega^2(r,\theta)}}, \label{12p}
\end{equation}
where $\omega(r,\theta)$ is the frequency of photon measured by a static observer outside the event horizon. The Hamilton's equations are defined as
\begin{equation}
\dot{x}^\mu=\partial_\tau x^\mu=\partial_{p_\mu}\mathcal{H}, \qquad \dot{p}_\mu=\partial_\tau p_\mu=-\partial_{x^\mu}\mathcal{H} \label{14p}
\end{equation}
with $\tau$ being an affine parameter. Since, a plasma medium is dense and dispersive, therefore, the motion of photons is affected by the frequencies of electrons in the plasma. Therefore, the frequency of the propagating photon must be greater than the frequency of the electrons in plasma, that is, $\omega^2(r,\theta)\geq\omega^2_\text{p}(r,\theta)$. The observer is assumed static with the four-velocity in comoving coordinates given as $U^\mu(r,\theta)=\left(-g_{tt}(r,\theta)\right)^{-1/2}$ and the photon frequency is $\omega(r,\theta)=-p_\mu U^\mu(r,\theta)=-p_t U^t(r,\theta)$. The Planck's relation connects the energy and the angular frequency as $E=\hbar\omega_0$ which gives $p_t=-\omega_0$ by setting the units for $\hbar=1$. Therefore, one obtains
\begin{equation}
\omega(r,\theta)=\omega_0\left(-g_{tt}(r,\theta)\right)^{-\frac{1}{2}}. \label{15p}
\end{equation}
We obtain $t$ and $\phi$ components of geodesic equations corresponding to two constants of motion, energy $E$ and angular momentum $L$. For other two equations, we consider Hamilton-Jacobi equation in the form
\begin{equation}
g^{\mu\nu}\partial_{x^\mu}\mathcal{S}\partial_{x^\nu}\mathcal{S}+\omega^2_\text{p}(r,\theta)=0 \label{18p}
\end{equation}
with Jacobi action given by Eq.~(\ref{jact}) with $m_\text{p}=0$ as the third constant of motion. For the given metric, the Hamilton-Jacobi equation
\begin{align}
&\Delta(r)\left(\partial_r\mathcal{A}_r(r)\right)^2+\left(L^2\csc^2\theta-a^2E^2\right)\cos^2\theta+(L-aE)^2\nonumber\\&+\left(\partial_\theta\mathcal{A}_\theta(\theta)\right)^2-\frac{\left(\left(r^2+a^2\right)E-aL\right)^2}{\Delta(r)}+\rho^2\omega_\text{p}^2(r,\theta)=0 \label{20p}
\end{align}
cannot be separated until we assume \cite{PhysRevD.95.104003}
\begin{equation}
\omega_\text{p}(r,\theta)=\frac{\sqrt{f_r(r)+f_\theta(\theta)}}{\rho}, \label{21p}
\end{equation}
where, $f_r(r)$ and $f_\theta(\theta)$ are arbitrary functions that ensures the separability of the Eq.~(\ref{20p}). Therefore, generating the Carter constant $\mathcal{Z}$ as fourth constant of motion, the null geodesic equations are of same form as Eqs.~(\ref{teq})-(\ref{pheq}) but with the functions
\begin{eqnarray}
\mathcal{R}(r)&=&\left(aL-\left(r^2+a^2\right)E\right)^2\nonumber\\&&-\Delta(r)\left(f_r(r)+\mathcal{Z}+(aE-L)^2\right), \label{28p} \\
\Theta(\theta)&=&\mathcal{Z}+a^2E^2\cos^2\theta-L^2\cot^2\theta-f_\theta(\theta). \label{29p}
\end{eqnarray}
The impact parameters $\xi$ and $\eta$ in presence of plasma become
\begin{align}
\xi(r_\text{p})&=\frac{1}{a\Delta'(r_\text{p})}\Big[\left(r_\text{p}^2+a^2\right)\Delta'(r_\text{p})-\Delta(r_\text{p})\Big(2r_\text{p}\nonumber\\&+\sqrt{4r_\text{p}^2-f_r'(r_\text{p})\Delta'(r_\text{p})}\Big)\Big], \label{62}\\
\eta(r_\text{p})&=\frac{1}{a^2\Delta'^2(r_\text{p})}\Big[8r_\text{p}^2\Delta(r_\text{p})\big[a^2-\Delta(r_\text{p})\big]-\big[r_\text{p}^4\nonumber\\&+a^2f_r(r_\text{p})\big]\Delta'^2(r_\text{p})+\big[4r_\text{p}^3-\big(a^2\nonumber\\&-\Delta(r_\text{p})\big)f_r'(r_\text{p})\big]\Delta(r_\text{p})\Delta'(r_\text{p})+2r_\text{p}\Delta(r_\text{p})\big[2\big(a^2\nonumber\\&-\Delta(r_\text{p})\big)+r_\text{p}\Delta'(r_\text{p})\big]\sqrt{4r_\text{p}^2-f_r'(r_\text{p})\Delta'(r_\text{p})}\Big]. \label{63}
\end{align}
Now considering two cases for the functions $f_r(r)$ and $f_\theta(\theta)$, we will construct celestial coordinates to investigate shadows. The \textbf{case I} corresponds to the assumption $f_r(r)=\omega_\text{c}^2\sqrt{M^3r}$ and $f_\theta(\theta)=0$, where, $\omega_c$ has the dimensions of frequency. The celestial coordinates for this case are of the same form as in Eqs.~(\ref{celx}) and (\ref{cely}) with their respective impact parameters given in Eqs.~(\ref{62}) and (\ref{63}). For the equatorial observer, the celestial coordinates reduce to the same form as in Eq.~(\ref{cele}). The \textbf{case II} corresponds to the assumption $f_r(r)=0$ and $f_\theta(\theta)=\omega_\text{c}^2M^2\left(1+2\sin^2\theta\right)$ with the same parameter $\omega_c$. The horizontal component of the celestial coordinates for this case is of the same form as in Eq.~(\ref{celx}), however, the vertical component is modified. For an observer at $\left(\infty,\theta_0\right)$, it becomes
\begin{figure}[t!]
\centering
\subfigure{\includegraphics[width=0.38\textwidth]{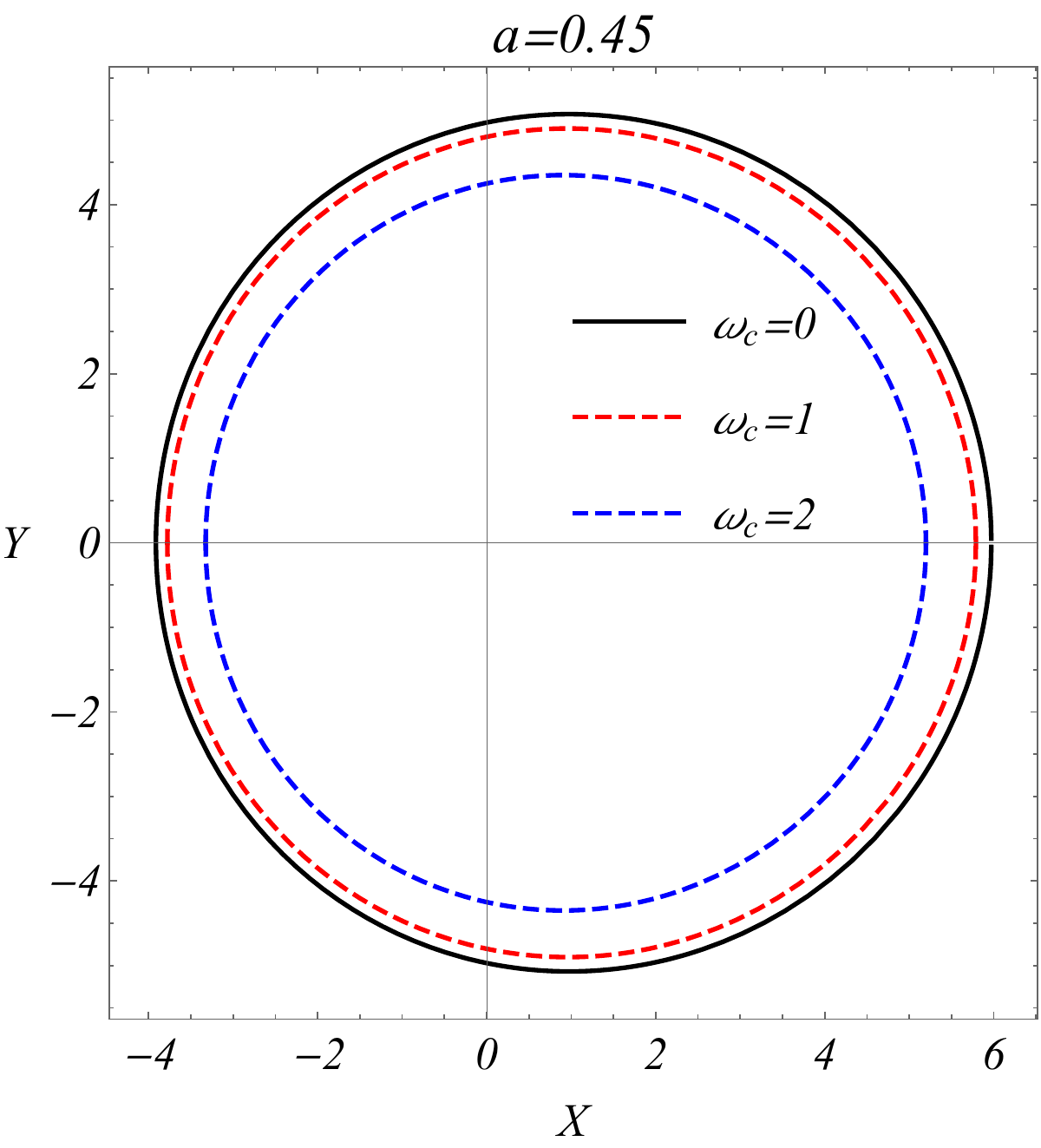}}
\subfigure{\includegraphics[width=0.38\textwidth]{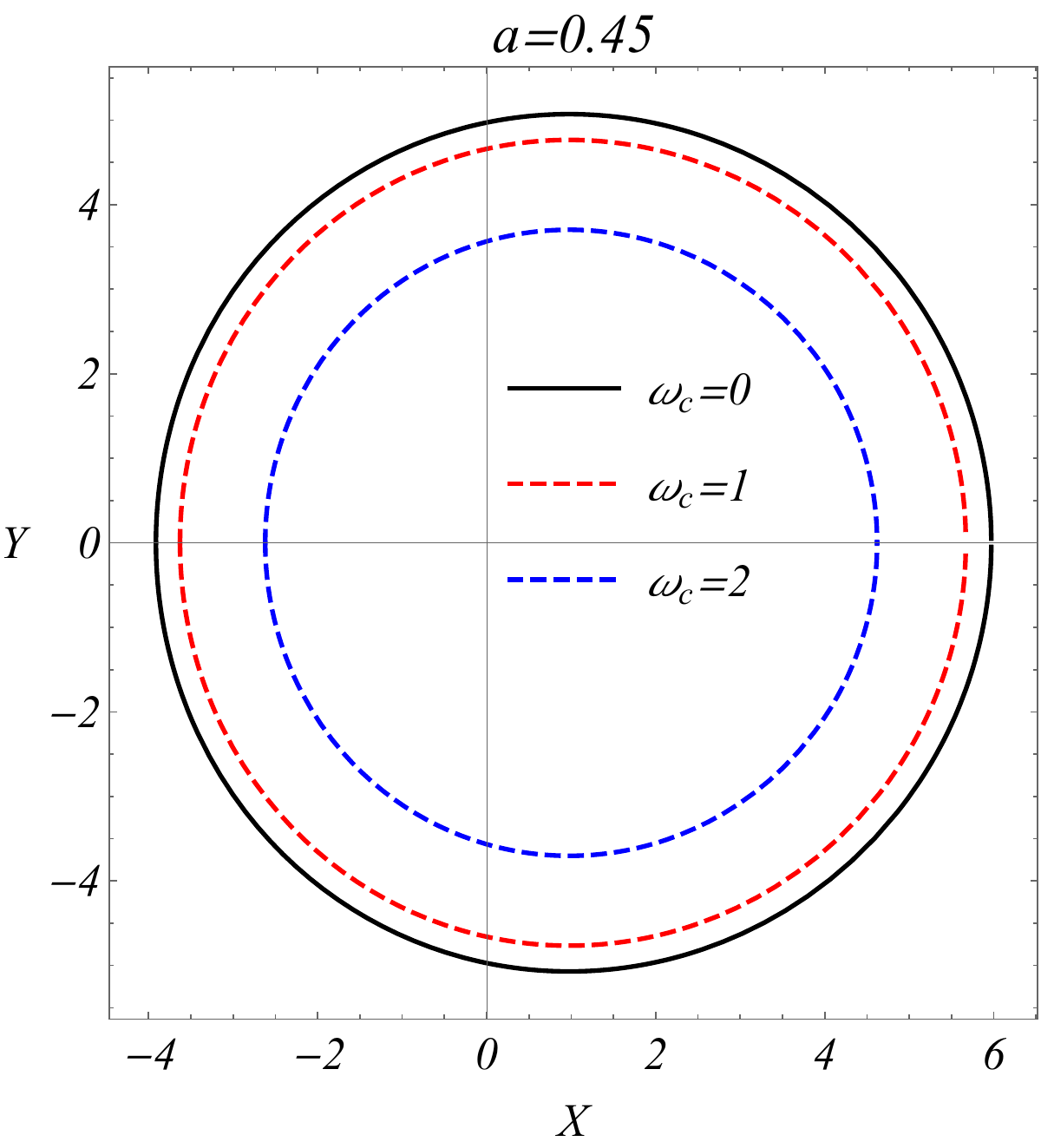}}
\caption{Influence of $\omega_\text{c}$ in case I (upper panel) and case II (lower panel) on shadows of quantum corrected Kerr BH for a fixed value of $a$. \label{F5}}
\end{figure}
\begin{eqnarray}
Y(r_\text{p})&=&\pm\Big[\eta(r_\text{p})+a^2\cos^2\theta_0-\xi^2(r_\text{p})\cot^2\theta_0\nonumber\\&&-\omega_\text{c}^2M^2\left(1+2\sin^2\theta_0\right)\Big]^{\frac{1}{2}} \label{69}
\end{eqnarray}
and for the equatorial observer, it reduces to
\begin{equation}
Y(r_\text{p})=\pm\sqrt{\eta(r_\text{p})-3\omega_\text{c}^2M^2}. \label{71}
\end{equation}
For both of above mentioned cases, we have plotted the shadows in Fig.~\ref{F5} for a few values of plasma parameter $\omega_\text{c}$ and keeping a fixed spin of the quantum corrected Kerr BH. The spin deviates the shadow from a pure circular loop causing a distortion. While, the plasma parameter $\omega_\text{c}$ reduces the size of the shadow. It can also be seen that $\omega_\text{c}$ in case II is more significant and has higher sensitivity as compared to the case I. The variation in the shadow size for case II is greater than the variation observed in case I.

\section{Discussion}\label{S5}
By a smooth matching of APS and qOS spacetimes by identifying a set of suitable coordinate transformation, a quantum corrected Schwarzschild BH in LQG is obtained with a relatively weak extra term of order $r^{-4}$ depending on Barbero-Immirzi parameter $\gamma$ \cite{PhysRevLett.130.101501}. During the collapse of the dust ball, corresponding to a lower bound of the radial coordinate, a lower bound on the mass of this BH is obtained, below which, there exists no horizon, whereas, a two horizon system is obtained for all mass values above the minimum mass limit. Ye et al. \cite{YE2024138566} studied the shadows of this BH to determine the effect of LQG on the BH. They considered the fixed value of the Barbero-Immirzi parameter and hence the parameter $\alpha$. They presented their findings in comparison with the results of Schwarzschild BH. Motivated by this, we considered the rotating counterpart of the quantum corrected Schwarzschild BH, in effective LQG and accomplished our analysis in comparison with the results for Kerr BH.

The rotating BH metric (\ref{romet}) being an effective metric still encompasses various features of Kerr-like BH, especially the existence of time translation and rotational invariance isometries. Additionally, on removing the spin, its exact static counterpart is recovered, and on removing the quantum effects, the rotating metric reduces to Kerr metric and the static metric reduces to Schwarzschild metric. The parameter $\alpha$ though appears as the LQG parameter in the BH metric, however, it does not behave as a free parameter and has a fixed value that cannot be varied. The metric function of the quantum corrected Kerr BH is expressed in terms of the metric function of Kerr BH for which the effect of LQG can be observed in the Eq.~(\ref{delK}) and from the Fig.~\ref{F1}. The quantum corrected Kerr BH in effective LQG also exhibits two horizons, though the event horizon being small as compared to the event horizon of Kerr BH, reduces the size of BH due to LQG effects. Its extreme spin value is also reduced to the half of extreme spin of Kerr BH.

The function $\mathcal{R}(r)$ plays a vital role in particle orbits and shadow analysis. Like the metric function, it is also expressed in terms of the function corresponding to the Kerr BH which generates an identical form of effective potential as well. This form enables us to generalize the location of unstable null orbits for such kind of metrics. We proved a theorem based on convexity and the familiar results of effective potential of Kerr BH, which ensures that the unstable orbits for such a quantum corrected Kerr BH will be smaller than the unstable orbits for Kerr BH. This result holds true only if the extra term in effective potential function is decreasing function of $r$. The result in this theorem is then verified numerically in the Fig.~\ref{F2}. The impact parameters and celestial coordinates for quantum corrected Kerr BH are also expressed in terms of the impact parameters and celestial coordinates for Kerr BH. The deviation terms in the celestial coordinates determine the deviation of shadow of quantum corrected Kerr BH from the shadow of Kerr BH. The LQG effects influences both the size of the shadow and the distortion in it. Since, the extreme spin for quantum corrected Kerr BH is $\sim$0.4952, so the ergosphere does not get enough strength to create a flat shadow like the shadow of Kerr BH.

The astrophysical impact of LQG on quantum corrected Kerr BH is investigated through the comparison of the shadow size for Kerr and quantum corrected Kerr BHs with the size of M87* and Sgr A*. Along with the images, the data from EHT results enabled us to draw an analysis to determine the constraints on the BH spin parameters and the influence of LQG on it. The Kerr BH is defined for $0\leq a\leq1$, however, it becomes identical with M87* for $0\leq a\leq0.586$. Whereas, the effect of LQG on quantum corrected Kerr BH inhibits it to behave like M87* for all spin values. On the other hand, both Kerr and quantum corrected Kerr BHs mimic Sgr A* for all spin values, however, due to LQG effects, the quantum corrected Kerr BH is more likely to mimic Sgr A* than Kerr BH.

The plasma surrounding the quantum corrected Kerr BH in effective LQG has a great impact on the light propagation. The shadow size is reduced by increasing the value of $\omega_\text{c}$ in both cases. However, the quantity of variation in shadow size reveals that $\omega_\text{c}$ in case II is more sensitive and has greater impact than that in case I. 

As a future project, one may investigate the effect of LQG on the deflection of light in strong and weak regimes of quantum corrected Kerr BH. The study of BH evaporation rate, Hawking radiation via tunneling process and Unruh effect would also be an intriguing analysis. Moreover, it will be interesting to study the effect of LQG on the quantum corrected Kerr BH behaving as a particle accelerator.

\section*{Acknowledgment}
This paper is dedicated to the memory of Prof. Jerzy Lewandowski, who derived the static quantum corrected BH metric in LQG which is considered in this paper. He is no more with us but his work will continue to inspire us, leaving a lasting legacy in theoretical physics. We are thankful to Francesco Fazzini for discussions and useful suggestions for improvements in the paper.

\bibliography{refs.bib}
\bibliographystyle{apsrev4-2}

\end{document}